\documentclass[twocolumn,superscriptaddress,amsmath,amssymb,prl]{revtex4}
\usepackage{graphicx}
\usepackage{dcolumn}
\usepackage{bm}

\newcommand{\sto}{SrTiO$_3$}
\newcommand{\lao}{LaAlO$_3$}
\newcommand{\dir}{$\langle001\rangle$}
\newcommand{\qo}{$q_{0}$}
\newcommand{\Lisl}{$L_{isl}$}
\newcommand{\Ispec}{$I_{spec}$}
\newcommand{\Iisl}{$I_{isl}$}
\newcommand{\tspec}{$\tau_{spec}$}
\newcommand{\tisl}{$\tau_{isl}$}

\begin{document}

\title{Thickness dependence of surface diffusion in epitaxial \lao\ on \sto\dir}

\author{J. D. Ferguson}
 \affiliation{Department of Materials Science and Engineering, Cornell University, Ithaca, NY 14853, USA}
 \affiliation{Cornell Center for Materials Research, Cornell University, Ithaca, NY 14853, USA}
\author{Y. Kim}
 \affiliation{School of Applied and Engineering Physics, Cornell University, Ithaca, NY 14853, USA}
 \affiliation{Cornell Center for Materials Research, Cornell University, Ithaca, NY 14853, USA}
\author{A. R. Woll}
\affiliation{Cornell High Energy Synchrotron Source, Cornell University, Ithaca, NY 14853, USA}
\author{J. D. Brock}
 \affiliation{School of Applied and Engineering Physics, Cornell University, Ithaca, NY 14853, USA}
 \affiliation{Cornell Center for Materials Research, Cornell University, Ithaca, NY 14853, USA}

\date{\today}

\begin{abstract}
The \lao/\sto\dir\ thin film materials system was studied using in situ, simultaneous x-ray diffuse scattering and specular reflectivity during pulsed laser deposition. Using this method, we are able to measure the time dependence of the characteristic surface length scale and the characteristic time for both in-plane and downhill diffusion. These data allow for the determination of the activation energy for various diffusion processes as a function of \lao\ thickness. Additionally, we show that the downhill diffusion rate of the first monolayer is distinctly different than subsequent layers. These results are directly compared to previous experimental observations seen during the deposition of homoepitaxial \sto\dir.
\end{abstract}


\maketitle

Complex oxides exhibit a vast range of materials properties encompasing electrical insulators, high-$T_{c}$ superconductors, semiconductors, dielectrics, ferromagnetics, and multiferroics. 
Of particular note, these materials admit the deterministic synthesis of essentially arbitrary crystalline structures, which is leading to new discoveries.
For example, for certain thicknesses,
the \lao/\sto\dir\ system exhibits
a conducting, quasi-two-dimensional electron gas at the interface between two wide-bandgap insulators \cite{Ohtomo:2004rw, Brinkman:2007hc, N.Reyren08312007, J.Mannhart03262010}.
The formation of the conducting layer is sensitive to the number of \lao\ layers \cite{S.Thiel09292006}, which leads to the possibility of new device architectures \cite{cen:2009dj}. 
Pulsed laser deposition (PLD), long known as the tool of choice in a research context for the growth of complex-oxide materials, has now demonstrated 
the ability to grow interfaces with a sharpness previously thought attainable only with molecular beam epitaxy.
With the keen interest in growing such heterostructures and artificial materials via PLD, it is critically important to understand the microscopic growth processes setting the fundamental limits to obtaining atomically perfect growth.
Here, time-resolved, simultaneous diffuse and specular x-ray scattering measurements are used to determine the rate of several surface diffusion processes during the PLD of \lao\ on \sto\dir. 
Perhaps not surprisingly, the phenomenology we observe in this heteroepitaxial system is much richer than that observed in homoepitaxy, with film thickness dependent kinetic coefficients.
We find that the barrier for in-plane diffusion of \lao\ on \sto\ is substantially larger than the rate-limiting barrier for downhill transport from an \lao\ island to the \sto\ substrate. 
And, the energy barrier for diffusion is found to be a function of \lao\ film thickness, for films from one to six unit cells thick.

These studies were performed in the PLD/x-ray diffraction chamber in the G3 experimental station of the Cornell High Energy Synchrotron Source (CHESS). 
The supplemental documents of Ref.\ \cite{Ferguson:2009} contain the experimental details. 
A one-dimensional diode array detector (pixel size $0.125\times5$ mm) was used to acquire the diffuse scattering data at the quarter-Bragg position on the crystal truncation rod.
A KrF excimer laser (248 nm), focused down to a 7.4 mm$^{2}$ spot, was used to ablate the single crystal \lao\ target. 
The laser fluence was 1.6 J cm$^{-2}$ for all films, producing a deposition flux of approximately twelve pulses-per-monolayer. 
Here, a monolayer (ML) refers to a one unit cell thick layer. 
The laser was fired at a repetition rate of 0.19 Hz. 
The substrate was held at controlled temperatue in a background pressure of $7.5\times10^{-6}$ Torr of O$_{2}$.
All films were grown in a layer-by-layer growth mode to a thickness of six MLs, which is well below the critical thickness for strain relaxation \cite{Merckling200747}.

In layer-by-layer growth, islands nucleate, grow, and then coalesce. 
At high coverage, near layer completion, the surface is best described by a series of pits that fill in as more material is deposited. 
During deposition, a characteristic length scale \Lisl, is present on the surface. 
This length scale is a measure of the average separation between islands (or pits) \cite{Evans20061}. 
At low coverage, x-ray diffuse scattering directly measures both $L_{isl}$ and the characteristic time for in-plane (intra-layer) mass transfer, $\tau_{isl}$ \cite{Ferguson:2009}. 
By monitoring the surface roughness using the specular reflectivity, the characteristic diffusion time for downhill (inter-layer) transport, $\tau_{spec}$, may be determined \cite{Blank:1999, Ferguson:2009, Fleet:2005, Lippmaa:2000, Tischler:2006}. 
Identifying 
the diffusion length as, $L_{d}=L_{isl}/2$, the intra- and inter-layer diffusion rates may be calculated by invoking the Einstein relation: $D=L_{d}^{2}/4\tau$ \cite{Gomer:1990}.


The x-ray scattering may be broken into specular \Ispec and  diffuse $I_{diff}$ components and each component can be accurately
described by the functional form
\begin{equation}
I(q_{||})=I_{0}/[1+\xi^{2}(q_{||}-q_{0})^{2}]^{3/2},
\label{eq1}
\end{equation}
where $q_{||}$ is the in-plane scattering vector, and \qo\ is the peak position \cite{Ferguson:2009}. 
For $I_{diff}$, \qo\ is directly related to the distance between islands or pits by the equation: $q_{0}\approx2\pi/L_{isl}$ \cite{Dulot:2003}. 
For \Ispec, $q_0 = 0$.
During homoepitaxial layer-by-layer deposition, \Ispec\ oscillates due to the continuous roughening and smoothening of the surface. Each local intensity maximum corresponds to the completion of $\approx$1 ML \cite{Ferguson:2009, Fleet:2005, Tischler:2006}. 
In heteroepitaxy, interference between beams reflected by the interface and the surface create an additional modulation.
These interference effects are known as ``Kiessig'' fringes, and their periodicity is dependent on the out-of-plane scattering vector, $q_{z}$ \cite{Kiessig:1931dz}. For our scattering geometry, the envelope of $I_{spec}$ has a periodicity of $\approx$4 MLs, and the roughness oscillations within this envelope have maxima corresponding to the completion of $\approx$1 ML.

To interpret the data, we write the total scattering in a $q_z$ plane of reciprocal space as
\begin{equation}
I_{tot} = I_{spec} + I_{isl} + I_{sm},
\label{eq1}
\end{equation}
where $I_{isl}$ is the observed diffuse scattering\footnote{$I_{isl}$ is obtained by integrating the fit to the diffuse line, Eqn. \ref{eq1}, over the x-y plane: $I_{isl}=2\pi \int_{0}^{\infty} q\,I_{diff}(q)\,dq$}, 
and $I_{sm}$ accounts for scattering by features smaller than our detection limit. 
(Since our maximum measurable value of $q_{||}$ is $q_{max}\approx0.2$ \AA$^{-1}$, we are insensitive to surface fluctuations with wavelengths smaller than 31\AA.)
Consequently, if the coverage is low and \Ispec\ is constant, a time dependent \Iisl\ results from  mass transfer between very small  islands to islands measurable by our experiment. 
Therefore, the time dependence of \Iisl\ provides a direct measure of the time scale for intra-layer mass transport \cite{Ferguson:2009}. 

\begin{figure}[t]
\includegraphics{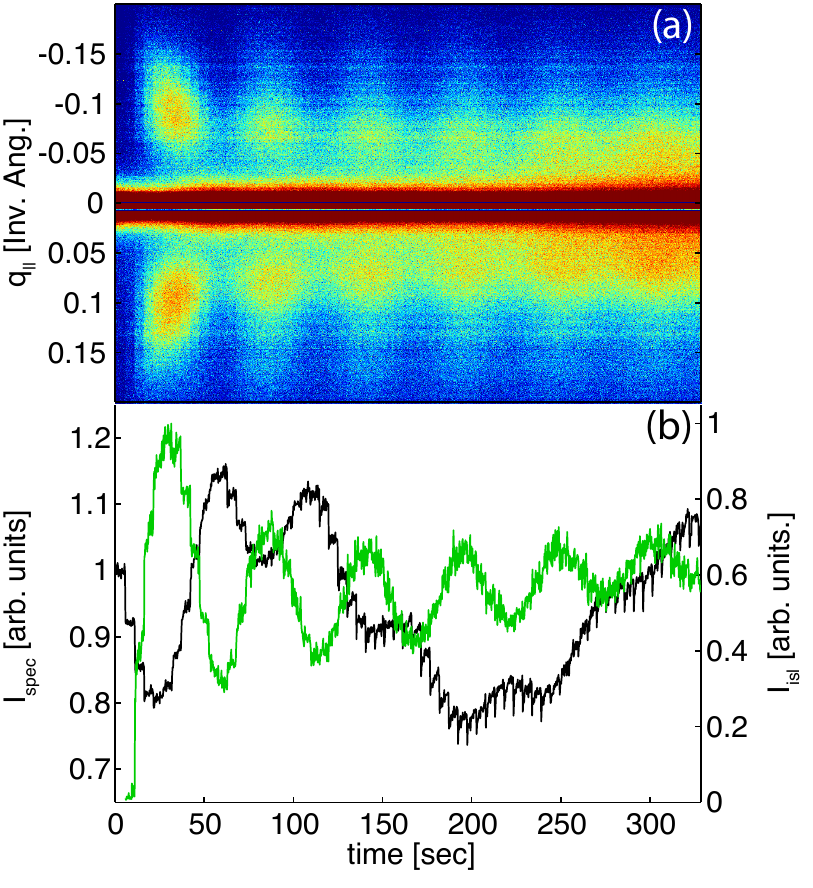}
\caption{\label{fig1} (a) X-ray scattering data for the PLD of \lao\ on \sto\dir\ at $781^{\circ}$C. The intensity measured at $q_{||}\neq0$ is the diffuse x-ray scattering. The diffuse scattering is directly related to the \lao\ island distribution. (b) The corresponding fit results for \Ispec\ (black) and \Iisl\ (green).}
\end{figure}

Figure \ref{fig1}(a) shows a false color image of the scattered x-ray intensity, as a function of time and $q_{||}$, during the deposition of six MLs of \lao\ on \sto. 
Each 260-msec time slice, represents intensity vs. $q_{||}$ (see Ref.\cite{Ferguson:2009} for details). 
The intensity seen at $q_{||}\neq0$ is a result of the Henzler ring of diffuse scattering \cite{Hahn:1980} and, as mentioned above, the size and shape of this ring is determined by the \lao\ island distribution on the surface \cite{Evans20061}. 
As observed during PLD of homoepitaxial \sto\ \cite{Ferguson:2009}, \qo\ decreases as each additional layer of the \lao\ film is deposited, representing a systematic decrease in the island density with increasing layer number. 
To decompose \Ispec\ and \Iisl, each time slice is fit to our model; the results are shown in Fig. \ref{fig1}(b). 
In this figure, \Iisl\ oscillates with the period of 1 ML, and is out of phase with \Ispec. 
As discussed above, both the roughness and Kiessig components to \Ispec\ are visible in  Fig. \ref{fig1}(b). 
The periodicity seen in \Ispec\ that is $\pi$ out of phase with \Iisl\  is the �roughness� component, and it results from the scattering of x-rays by the \lao\ islands \cite{Sinha:1996}. The minimum in the envelope of \Ispec\ at around 200 seconds is a signature of the �Kiessig� component, therefore, the $^{1}/$$_{4}$ Bragg position sets a 4:1 ratio between the diffuse and Kiessig periods.

\begin{figure}[t]
\includegraphics{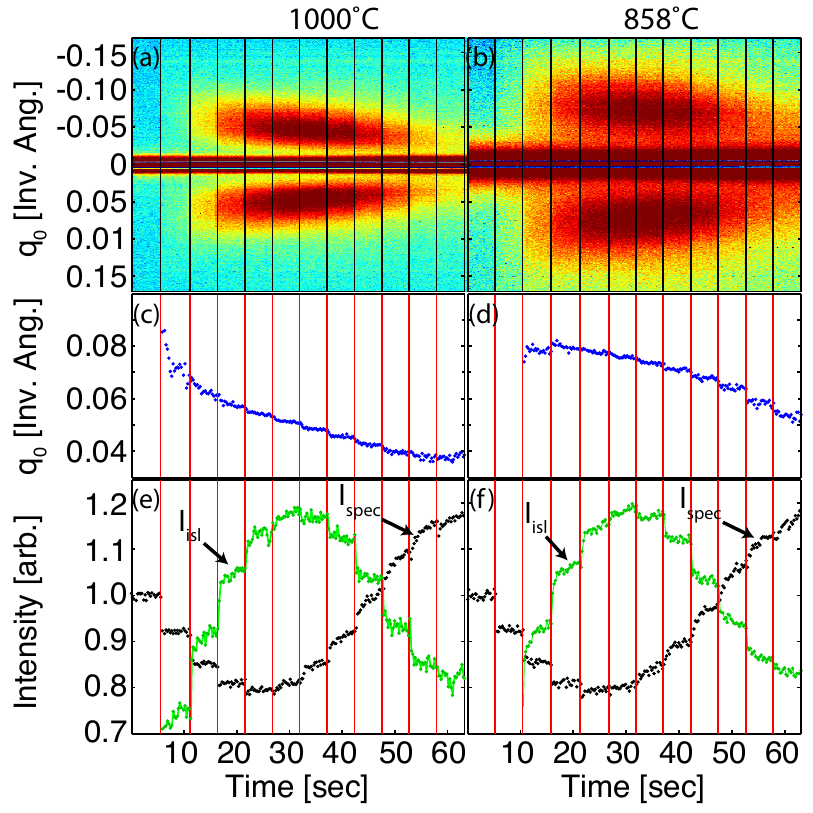}
\caption{\label{fig2} Diffuse scattering data and the corresponding fit results for the first ML of \lao\ at 1000$^{\circ}$C and 858$^{\circ}$C. The vertical lines represent the times when the laser was fired. (a), (c) and (e) correspond to the 1000$^{\circ}$C growth and (b), (d) and (f) show the 858$^{\circ}$C results. The arrows annotate the data points used to determine \tisl\ and \tspec.}
\end{figure}

To examine the influence of temperature on the \lao\ deposition, we show false color images of the scattering data obtained during the deposition of the first ML at 1000$^{\circ}$C and 858$^{\circ}$C in Figs. \ref{fig2}(a) and \ref{fig2}(b), respectively. Immediately noticeable in the data is the increase in \qo\ as the surface temperature is decreased. This implies that the island density increaseds with decreasing temperature, consistent with classical nucleation theory. These data were fit to our model, and the results are shown in Figs. \ref{fig2}(c)-\ref{fig2}(f). As seen in Fig. \ref{fig2}(c), \qo\ monotonically decreases both inter- and intra- pulse at 1000$^{\circ}$C. This behavior is attributed to coarsening of the \lao\ island distribution on the surface \cite{Bartelt:1996as, Evans20061, Ferguson:2009}. Fig. \ref{fig2}(d) shows \qo\ for the 858$^{\circ}$C deposition. At, and below, this temperature, the low diffuse intensity precluded an accurate fit before the second laser pulse. An interesting feature of Fig. \ref{fig2}(d) is the slight increase in \qo\ just after the third laser pulse. The best fit estimates for the peak positions just before and after the third laser pulses are $q_{0}=0.078\pm0.001$\AA$^{-1}$ and $q_{0}=0.081\pm0.001$\AA$^{-1}$, respectively. This increase in $q_{0}$, as additional material arrives indicates an increase in island density. Therefore, island nucleation may be occurring up to this coverage at 858$^{\circ}$C. 

Figures \ref{fig2}(e) and \ref{fig2}(f) show \Ispec\ and \Iisl\ for the data presented in Figs. \ref{fig2}(a) and \ref{fig2}(b), respectively. Following the third laser pulse, a constant \Ispec\ accompanies the slow rise in \Iisl. As discussed previously, this temporal dependence of \Iisl\ corresponds to the timescale for in-plane mass transfer. Additionally, the time dependence of \Ispec\ at high coverage is a direct measurement of the timescale for downhill diffusion. Fitting both \Iisl\ and \Ispec\ to a simple exponential function allows for the determination of these characteristic diffusion times.

\begin{figure}[b]
\includegraphics{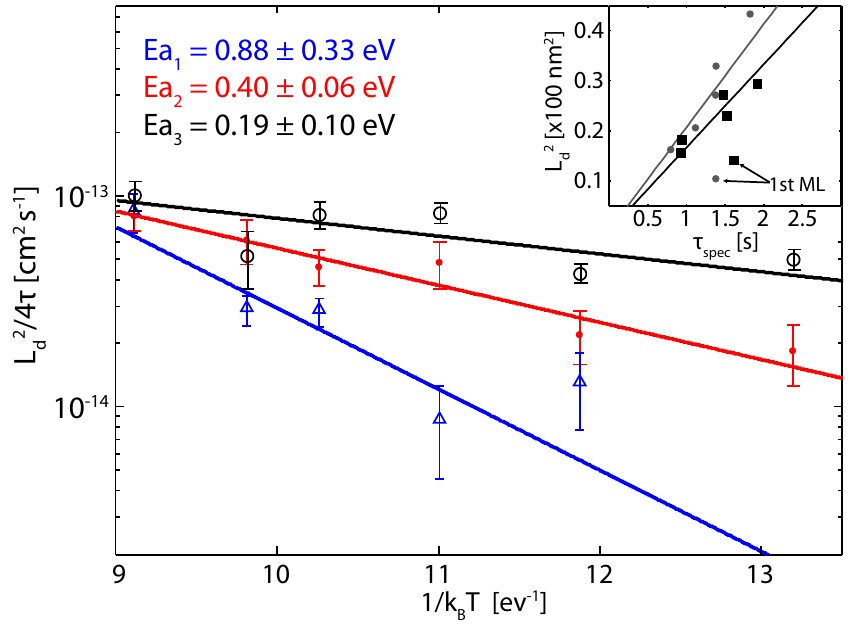}
\caption{\label{fig3} Arrhenius plot of the diffusion rate for \lao\ species during three different processes: in-plane diffusion on the \sto\ substrate (blue, triangles), downhill diffusion from the first \lao\ ML to the substrate (red, points), and downhill diffusion from a \lao\ ML to the \lao\ film (black, circles). The inset shows the dependence of $L_{d}$ on \tspec\ for each ML at 704$^{\circ}$C (squares) and 606$^{\circ}$C (circles). The arrows point to the data points for the first MLs.}
\end{figure}

Conceptually, the deposition of the first layer in heteroepitaxy is distinctly different than deposition of subsequent layers. For example, the diffusion rate for \lao\ species on the \sto\ substrate need not be the same as diffusing \lao\ species on the \lao\ film. Additionally, a mobile species diffusing on top of the first \lao\ ML might interact with the \sto\ substrate during downhill diffusion. Therefore, one might expect a film thickness dependence of the diffusion rate until the \lao\ film reaches a critical thickness for substrate interaction effects. To examine this possibility, the inset to Fig. \ref{fig3} shows a plot of $L_{d}^{2}$ vs. \tspec\ for each of the 6 ML deposited at two substrate temperatures: 606$^{\circ}$C and 704$^{\circ}$C. We note that after the first ML, $L_{d}^{2}$ and \tspec\ show a linear relationship, where the slope of the line is related to the diffusion rate by: $L_{d}^{2}=4D\tau_{spec}$. The fact that the first data point is inconsistent with the remaining points suggests that the interlayer transport for the first layer is distinctly different than that of subsequent layers.

To further examine the diffusion behavior in the \lao\ films, an Arrhenius plot is shown in Fig. \ref{fig3} for three data sets: the intra-layer diffusion rate for the first ML, the inter-layer diffusion rate for the first ML, and the average of the inter-layer diffusion rates in the second through sixth MLs. The intra-layer diffusion barrier for \lao\ species on \sto\ was determined to be $E_{a1}=0.88\pm0.33$ eV. The noise in \Iisl\ precluded the determination of \tisl\ for subsequent MLs, therefore, the intra-layer diffusion rate for \lao\ species on \lao\ could not be measured. Fitting the inter-layer diffusion data to an Arrhenius model yielded activation energies of $E_{a2}=0.40\pm0.06$ eV and $E_{a3}=0.19\pm0.10$ eV for the first ML and second-sixth ML, respectively. These data illustrate the layer dependence of the activation barrier, showing that the largest energy barrier corresponds to the diffusion of \lao\ species on the \sto\ substrate.
\begin{figure}[t]
\includegraphics{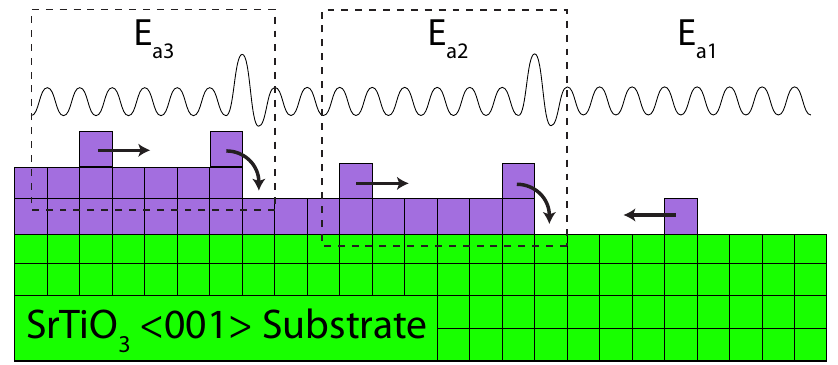}
\caption{\label{fig4} Schematic description of the activation barriers measured in Fig. \ref{fig3}. The oscillations show the various activation energies across the surface, with the large increase at the step edge representing the ES barrier. $E_{a1}$ is the in-plane diffusion barrier for \lao\ species on the substrate. $E_{a2}$ is the energy barrier for \lao\ species diffusing down to the \sto\ substrate. $E_{a3}$ represents the downhill activation barrier for layers 2-6. The dotted outline illustrates which set of diffusion processes determine $E_{a2}$ and $E_{a3}$.}
\end{figure}

The various diffusion processes measured from our data are presented schematically in Fig. \ref{fig4}. The diffusion barrier, $E_{a1}$, corresponds to in-plane mass transfer of \lao\ species on the \sto\ substrate. The activation barrier, $E_{a2}$, corresponds to the inter-layer diffusion from the top of the first \lao\ layer down to the substrate. $E_{a3}$ corresponds to the downhill diffusion of the second-sixth MLs to the underlying \lao\ layer. Our measurements do not allow us to determine if the inter-layer activation barriers are the Ehrlich-Schwoebel (ES) barrier, the surface diffusion barrier, or the sum of the two. 

Our results explicitly demonstrate the contrast between heteroepitaxial and homoepitaxial diffusion processes. Specifically, previous work measuring diffusion rates for PLD of homoepitaxial \sto\ showed that the energy barriers for inter- and intra-layer diffusion were the same: $E_{a}=0.97\pm0.07$ eV  \cite{Ferguson:2009}. It was also shown that the inter-layer diffusion rate is the same for all MLs during \sto\ homoepitaxial deposition (see Fig. 3 of ref. \cite{Ferguson:2009}). Here, the measured intra-layer diffusion barrier of the first \lao\ ML is found to be larger than the downhill activation barrier. Additionally, the inter-layer diffusion of the first ML of \lao\ on \sto\ was found to differ from subsequent layers, illustrating a distinct difference between homoepitaxy and heteroepitaxy.

Our data may be used to explain the persistent layer-by-layer growth mode for the \lao/\sto\ depositions. To do this, it is useful to consider the Volmer-Weber (3D) growth mode. In Volmer-Weber growth, the lower diffusion barrier of the film/substrate interface, when compared to the film/film diffusion barrier, results in the sticking of adatoms to the top of nucleated islands. This behavior has been seen in Fe$_{3}$Si growth on GaAs\dir, where the growth mode was directly correlated to the surface diffusion barriers \cite{kaganer:2009}. For \lao/\sto, $E_{a1}>E_{a2}$, providing the proper condition for inter-layer transport, a prerequisite for layer-by-layer growth. Therefore, diffusing species will tend to absorb to the substrate rather than the film. Additionally, this physical description may be applied to the deposition of the second ML of \lao, since $E_{a2}>E_{a3}$.

In conclusion, we have shown that simultaneous, time-resolved surface diffuse and specular scattering can be used to measure the surface diffusivity during the pulsed laser deposition of \lao\ on \sto\dir. We have used this data to measure energy barriers for three different surface diffusion processes: in-plane diffusion of \lao\ on \sto, downhill diffusion of \lao\ to the \sto\ substrate, and downhill diffusion of \lao\ to the \lao\ film. The activation barriers were found to decrease, respectively, for each of these processes, which was used to explain the persistent layer-by-layer growth mode. Additionally, we have shown that all diffusion barriers for \lao\ on \sto\ are small compared to homoepitaxial \sto.

\begin{acknowledgments}
We would like to thank D. Muller, D. Schlom, and D. Dale for useful conversations. This work was supported in part by the National Science Foundation (NSF) (DMR-0705361) and in part by the Cornell Center for Materials Research (CCMR) with funding from the Materials Research Science and Engineering Center program of the National Science Foundation (cooperative agreement DMR-0520404) and is based upon research conducted at the Cornell High Energy Synchrotron Source which is supported by the NSF and the National Institutes of Health/National Institute of General Medical Sciences under NSF award DMR-0225180.

\end{acknowledgments}

\bibliography{refs}
\end{document}